\documentclass[prl,aps,twocolumn,showpacs,amsmath,amssymb,floatfix,citeautoscript]{revtex4}
\usepackage{graphicx}
\usepackage{dcolumn}
\usepackage{bm}
\usepackage{amsmath}
\usepackage{amssymb}
\usepackage[normal]{subfigure}

\makeatletter

\usepackage{color}

\begin{document}

\title{Proposed Method for Distinguishing Majorana Peak from Other
    Peaks: Tunneling Spectroscopy with Ohmic Dissipation using Resistive Electrodes}

\author{Dong E. Liu}
\affiliation{
Department of Physics and Astronomy, Michigan State University, East Lansing, Michigan 48824, USA}

\date{\today}

\begin{abstract}
We propose a scheme to distinguish zero-energy peaks due to Majorana from those 
due to other effects at finite temperature by simply replacing the normal metallic lead 
with a resistive lead (large $R\sim k\Omega$) in the tunneling spectroscopy.
The dissipation effects due to the large resistance change the tunneling conductance
significantly in different ways. The Majorana peak remains increase as temperature 
decreases $G\sim T^{2r-1}$ for $r=e^2 R/h<1/2$.
The zero-energy peak due to other effects splits into two peaks at finite temperature
and the conductance at zero voltage bias varies with temperature by a power law.
The dissipative tunneling with a Majorana mode belongs to a same universal class
as the unstable critical point of the case with a non-Majorana mode.
\end{abstract}


\maketitle

\textit{Introduction} ---
Majorana fermions (MFs), proposed to exist in solid state systems~\cite{fu08,sau10,aliceaPRB10,Lutchyn10,oreg10},
cold atomic systems~\cite{sato09,zhu11,jiang11}, and periodic driving systems \cite{jiang11,reynoso12,liuPRL13},
attract a great deal of attention. A variety of signatures 
\cite{DasSarma06,fu&kane09,fu&kanePRB09,akhmerov09,law09,akhmerov11,wimmer11,liu11,jiangUCJE11,Fidkowski12,SanJose12} are predicted
to detect Majorana fermion (MF) zero mode;
among them, tunneling spectroscopy may provide one of the simplest and direct tests for MF---
The observation of the zero-bias peak (ZBP) with quantized conductance $G=2e^2/h$ \cite{law09,akhmerov11}
at sufficiently low temperature (smaller than intrinsic width of the Majorana peak). Recently, 
several groups \cite{kouwenhovenSCI12,deng12,das12} reported the observation of 
a non-quantized ZBP at higher temperature in semiconductor
nanowires, which is possibly coming from MF. 
However, the ZBP may originate from other effects, e.g. zero-energy impurity bound state. 
In addition, recent works \cite{Bagrets12,Liujie12,Neven13} show that, 
in a superconducting system with both spin-rotation and time-reversal symmetry breaking,
the disorder can induce a cluster of mid-gap states around zero-energy and
thus a ZBP at finite temperature. Especially, the disorder ZBP 
appears in the conditions highly similar to Majorana ZBP  \cite{Bagrets12,Liujie12,Neven13}.
These alternative possibilities lead to debates about the validity of the tunneling spectroscopy methods. 

In this work, we introduce a scheme by simply replacing the normal metal lead in the tunneling spectroscopy 
with a resistive lead (with large resistance $R\sim k\Omega$). In this case, electrons couple to 
an ohmic environmental bath \cite{Feynman&Vernon}
in the tunneling process; the coupling to the bath usually suppresses 
the tunneling rate and leads to dissipative tunneling \cite{LeggettRMP87,Nazarov&Ingold}. 
Dissipation effects can also cause non-trivial phase diagrams and transitions,
which was recently observed in a simple resonant level system \cite{Mebrahtu12,Mebrahtu13,LiuDRL13}.
We investigate how the dissipation influences the tunneling into MFs,
zero-energy impurity bound states in superconductor, and other states causing ZBP at finite temperature.
The ways that the dissipation effects renormalize the tunneling strength and the tunneling conductance is
significantly different for MFs and other cases. If the lead is connected to a MF,
the zero-bias conductance scales as $G\sim T^{2r-1}$ near a weak tunneling fixed point (high $T$) and
will go to perfect transmission $G=2e^2/h$ at $T=0$ for $r=e^2 R/h<1/2$.
If the lead is connected to a superconductor (SC) with a zero-energy impurity bound state (non-MF),
the system can be divided into four stable phases and an unstable symmetric point (i.e. critical point).
Away from the symmetric point, the system will flow to one of the four stable fixed points, near which the 
zero-bias conductance scales as $G\sim T^{2r}$ and the peak splits into two at finite temperature.
The critical point belongs to the same universal class as the case for dissipative tunneling into a Majorana mode.
We also consider the conductance for the dissipative tunneling 
into a cluster of mid-gap states.
Without dissipation, the finite temperature conductance shows 
ZBP; with dissipation, the single peak splits into two as temperature decreases. 
The splitting occurs at higher temperature for
larger resistance, but $r<1/2$ is required in the experiment so that Majorana ZBP does not split.
Therefore, the dissipation effect induced by the resistive lead provides a way to distinguish Majorana ZBP
and other ZBP, and serves as a `` Majorana signature filter''.

\begin{figure}[htp]
\centering
\includegraphics[width=3.0in,clip]{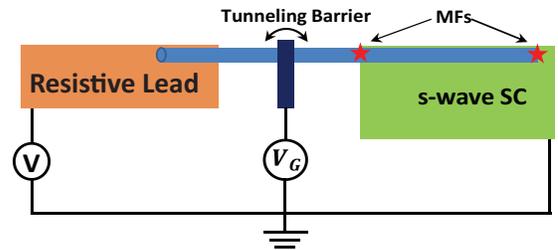}
\vspace{-0.1in} 
\caption{(color online) Proposed experimental setup. }
\vspace{-0.1in} 
\label{fig:setup}
\end{figure}

\textit{Model} --- We consider the tunneling spectroscopy from a resistive 
lead into the end of a superconducting nanowire (SCNW)
with Rashba spin-orbit coupling and proximity induced superconductivity $\Delta$ as shown
in Fig.~\ref{fig:setup}. A magnetic field is applied
perpendicular to the direction of the Rashba spin-orbit coupling. 
In this case, MFs are predicted to exist at the
two ends of the wire if $V_z>\sqrt{\Delta^2+\mu^2}$, 
where $V_z$ is Zeeman splitting and $\mu$ is wire chemical 
potential \cite{Lutchyn10,oreg10}. Unlike conventional setup, 
we replace the normal metallic lead with a resistive lead. 
A gate is applied to 
control the tunneling barrier between the lead and SCNW. We assume that the barrier
is high and wide, so that the tunneling has only a single channel, and 
the cooper pair tunneling can be assisted only by the mid-gap states
localized near the end of the wire. Note that our setup is not
limited only to SC wire, but also any other MF setups with a resistive lead.

The Hamiltonian of the system can be written as
\begin{equation}
 H = \sum_{k} (\epsilon_{k}+\mu_1) c^{\dagger}_{k} c_{k} + H_{\rm SCNW} + H_{\rm T} + H_{\rm ENV}, 
 \label{eq:HHH}
\end{equation}
where the first term describes the lead, with the electron creation (annihilation) operator $c^{\dagger}_{k}$ ($c_{k}$) . 
The second term represents the states near the end of the nanowire:
\begin{eqnarray}
 H_{\rm SCNW} &=& \sum_{\nu} (\varepsilon_{\nu}+\mu_2) b^{\dagger}_{\nu} b_{\nu} + \text{SC Pairing} + \text{Disorder} \nonumber\\
 &=& \mu_2 N_{\rm SCNW} + \sum_{q} \xi_q \gamma^{\dagger}_{q} \gamma_{q},
\end{eqnarray}
where $b^{\dagger}$ ($b$) is the creation (annihilation) operator for electrons. 
Including the cooper pairing terms and disorders,
one can diagonalize the Hamiltonian and reach the bogoliubov quasi-particle states $\gamma_q$, which
includes the MF and the disorder induced mid-gap states. 
$\mu_1$ and $\mu_2$ are chemical potentials for the lead and superconductor, respectively.
The voltage bias is $V=\mu_1-\mu_2$.
The tunneling Hamiltonian in the presence of dissipation \cite{Nazarov&Ingold} is
\begin{equation}
 H_{\rm T} = \sum_{k,\nu} \Big( y_{k,\nu} c^{\dagger}_{k} b_{\nu} e^{-i \phi} + y_{k,\nu}^{*} b^{\dagger}_{\nu} c_{k} e^{i \phi} \Big),
 \label{eq:tunnelingH}
\end{equation}
where $y_{k,\nu}$ is the tunneling strength between lead and SCNW. The operator $\phi=(e/h)\int_{-\infty}^{t} dt' U(t')$ represents 
the phase fluctuation across the tunneling junction, where $U(t)$ is the voltage fluctuation across the junction. 
Define $Q$ as the charge fluctuation of the junction capacitance such that $[\phi, Q]=i\,e$. 
The operator $e^{-i \phi}$ removes one electron from the junction capacitance, and thus represents the single electron tunneling.
Following Caldeira and Leggett \cite{caldeira81}, one can represent the dissipative environment
by a set of harmonic oscillators (i.e. $\{q_n, \phi_n\}$ with oscillator frequency $\omega_n=1/\sqrt{L_n C_n}$)
bilinearly coupled to the phase $\phi$. The last term of Eq.~(\ref{eq:HHH})
is then \cite{caldeira81,LeggettRMP87,Nazarov&Ingold}
\begin{equation}
 H_{\rm ENV} = \frac{Q^2}{2 C} + \sum_{n=1}^{N} \Big[ \frac{q_n^2}{2 C_n} +\big(\frac{\hbar}{e}\big)^2 \frac{1}{2 L_n}(\phi-\phi_n)^2 \Big],
\end{equation}
where $C$ is the capacitance of the junction.
$H_{\rm ENV}$ describes the coupling between the system and the environment. 

\textit{Tunneling into Majorana Fermion} --- 
Consider the tunneling between the lead
and a MF zero-energy state, one arrives at the following Hamiltonian
\begin{equation}
 H_{\rm T} = \sum_{k} \Big( y_{k} c^{\dagger}_{k} \gamma_1 e^{-i \phi} + y_{k}^{*} \gamma_1 c_{k} e^{i \phi} \Big),
\end{equation}
where $\gamma_1=\gamma_1^{\dagger}$ is the MF operator. 
Note that, even for a spinful lead, MF couples to only a single channel, 
which is the linear combination of the spin up and down channels \cite{law09}. 
It is helpful to introduce a Dirac fermion $f$:
$\gamma_1 = (f+f^{\dagger})/\sqrt{2}$. The tunneling Hamiltonian becomes
\begin{eqnarray}
 H_{\rm T} &=& \frac{1}{\sqrt{2}} \sum_{k} \Big( y_{k} c^{\dagger}_{k} f e^{-i \phi} + y_{k}^{*} f^{\dagger} c_{k} e^{i \phi} \Big)  \nonumber\\
 & & + \frac{1}{\sqrt{2}} \sum_{k} \Big( y_{k} c^{\dagger}_{k} f^{\dagger} e^{-i \phi} + y_{k}^{*} f c_{k} e^{i \phi} \Big).
 \label{eq:tunneling_MF}
\end{eqnarray}
Now, a scaling analysis is in order to see how the tunneling strength $y$ scales in the renormalization group (RG) picture.
Because MF couples to the lead at a single point, the metallic lead can be reduced to a semi-infinite one dimensional 
free fermion bath \cite{HewsonBook}. Therefore, the scaling dimension of this fermion operator is $[c]=1/2$. The localized MF operator or
operator $f$ does not contribute to the scaling dimension. To study the phase part $e^{-i \phi}$, we consider an ideal ohmic dissipative
environment with the lead resistance $R$. If we are interested in the scaling dimension,
one only need the $T=0$ correlation function in the long time
limit $\langle e^{i \phi (t)}  e^{-i \phi (0)} \rangle
  \sim t^{-2 r}$ \cite{Nazarov&Ingold},
where $r=R/R_K$ with quantum resistance $R_K=h/e^2$.
We choose $\hbar=k_{B}=1$ throughout the paper. 
Therefore, the scaling dimension of the dissipative part is $[e^{-i \phi}]=r$, 
and the RG equation for the tunneling strength yields
\begin{equation}
 \frac{d y}{d \ln l} = \big( 1 - \frac{1}{2} - r  \big) y,
\end{equation}
where $l$ is a time cutoff.
For very large resistance $r>1/2$, the tunneling is an irrelevant perturbation and will flow to zero at zero energy.
However, for  $r<1/2$, the tunneling is relevant and will increase with reducing energy.
Near a weak tunneling fixed point (large $V$ or $T$) , the conductance scales as $G \sim V^{-2(1-1/2-r)}=V^{2r-1}$ at $T=0$,
and as $G \sim T^{2r-1}$ at $V=0$. As energy (i.e. $\rm{max}[V,T]$) approaches zero, the system will enter into a perfect
transmission case with quantum conductance $G=2e^2/h$ \cite{law09}.

\begin{figure}[t]
\centering
\vspace{0.1in} 
\includegraphics[width=2.4in,clip]{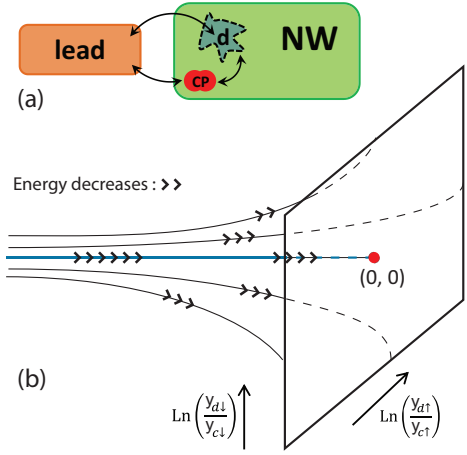}
\caption{(color online) (a) Demonstration of tunneling into a Zero-energy impurity bound 
states (non-Majorana). $d$ and $CP$ represent ZEIBS and cooper pair, respectively.
(b) Schematic representation of the flow diagram based on Eq. ~(\ref{eq:RGEQ_spinful}).
The arrows indicate the direction of the flow as energy decreases. The red dot in the center is the symmetric
fixed point ($y_{d\uparrow}=y_{c\uparrow}$ and  $y_{d\downarrow}=y_{c\downarrow}$), which is unstable.
The edges of the parallelogram correspond to four stable fixed points. 1) ($y_{d\uparrow}$ perfect transmission,
$y_{c\uparrow}=y_{d\downarrow}=y_{c\downarrow}=0$) at right edge. Note that $y_{d\downarrow}$
and  $y_{c\downarrow}$ have the same power law decay rate, and therefore $\ln(y_{d\downarrow}/y_{c\downarrow})=\rm{constant}$ near the Fixed point.
Other three fixed points are: 2) ($y_{c\uparrow}$ perfect transmission,
$y_{d\uparrow}=y_{d\downarrow}=y_{c\downarrow}=0$); 3) ($y_{d\downarrow}$ perfect transmission,
$y_{c\uparrow}=y_{d\uparrow}=y_{c\downarrow}=0$); 4) ($y_{c\downarrow}$ perfect transmission,
$y_{c\uparrow}=y_{d\uparrow}=y_{d\downarrow}=0$). 
}
\label{fig:impurityBS}
\end{figure}

\textit{Tunneling into Zero-Energy impurity Bound States (ZEIBS)} --- 
We assume a (non-MF) ZEIBS localized near the 
end of the wire as shown in Fig.~\ref{fig:impurityBS} (a). 
Suppose the ZEIBS and SC states consist of both spin up and down components, both spin channels in the lead
couple to them. 
These tunneling processes can be categorized as two mechanisms shown in Fig.~\ref{fig:impurityBS}: 1)
direct tunneling between the lead and ZEIBS, 2) tunneling into SC assisted by ZEIBS-SC tunneling with 
a cooper pair. The corresponding Hamiltonian is
\begin{equation}
 H_{T} = \sum_{\sigma} y_{\rm d, \sigma} \Psi_{\rm L, \sigma}^{\dagger}(0)\, d\, e^{-i\phi} + y_{\rm c, \sigma} \Psi_{\rm L, \sigma}^{\dagger}(0) d^{\dagger}  e^{-i\phi} e^{-i\chi}  + h.c.,
\end{equation}
where $y_{d, \sigma}$ and $y_{c, \sigma}$ are the tunneling strength for the lead-ZEIBS and lead-SC continuum ($\sigma$ represents the spin), 
$\Psi_{\rm L, \sigma}(0)=\sum_{k} \psi_{k, \sigma}(0) c_{k, \sigma}$ is the electron annihilation
operator of the lead at the point ($x=0$) coupled to SCNW, where $\psi_{k}$ is the wavefunction amplitude for state $k$. 
$\chi$ is the superconducting phase, and $e^{\pm i\chi}$ creates
or annihilates a cooper pair. We assume the SCNW is large enough to neglect the Coulomb charging energy, and the 
superconducting phase does not couple to any dissipative environment.  Under these assumptions, we can neglect the superconducting phase $\chi$, 
and then, the tunneling Hamiltonian is equivalent to the case with MF shown in Eq.~(\ref{eq:tunneling_MF}) 
if and only if $y_{d,\sigma}=y_{c,\sigma}$. 

Since the tunneling has only a single channel, the lead can be reduced to a semi-infinite free fermion field, which then can be unfolded to form
a chiral free fermionic field \cite{affleck95}; we take the coupling to the SCNW to be $x=0$. 
Then, this field can be bosonized in a standard way \cite{Senechal,Giamarchi}:
$\Psi_{\rm L \sigma}(x) = F_{\sigma}\, e^{i \Phi_{\sigma}(x)}/\sqrt{2\pi}$,
where $\Phi_{\sigma}(x)$ is a chiral bosonic field with
$[\Phi_{\sigma}(x),\Phi_{\sigma'}(x')]= i\delta_{\sigma \sigma'} \pi\, \rm{sgn} (x-x')$, $F_{\sigma}$ 
is Klein factor. For a spinful lead, 
the Hamiltonian becomes
\begin{eqnarray}
 H &=&  \sum_{\sigma}  \frac{v_F}{4\pi} \int_{-\infty}^{\infty} dx \big( \partial_{x}\Phi_{\sigma}(x) \big)^2 \nonumber\\
 && + \Big[ y_{\rm d, \sigma}  \frac{F_{\sigma}  e^{-i \Phi_{\sigma}(0)}}{\sqrt{2\pi}}  d\, e^{-i\phi}
  + y_{\rm c,\sigma} \frac{F_{\sigma} e^{-i \Phi_{\sigma}(0)}}{\sqrt{2\pi}}  d^{\dagger} e^{-i\phi}   \nonumber\\
  &&+ h.c. \Big]
  + K_{\sigma}(d^{\dagger}d-1/2)\partial_{x}\Phi_{\sigma}(x=0)/\pi.
\end{eqnarray}
The last term represents the density interaction between the lead 
(i.e. $\Psi_{\rm L \sigma}^{\dagger}(x)\Psi_{\rm L \sigma}(x)=-\partial_{x}\Phi_{\sigma}(x)/\pi$) 
and the localized ZEIBS, and this interaction is initially very small and can be
enhanced in the RG processes. Since the correlation function of the phase $\phi$ shows the similar power law decay
to the chiral bosonic field : $\langle e^{-i\phi(t)} e^{i\phi(0)} \rangle\sim t^{-2 r}$ and 
$\langle e^{-i\Phi_{\sigma}(x=0,t)} e^{i\Phi_{\sigma}(x=0,0)} \rangle\sim t^{-1}$,
we can combine the two bosonic field and introduce a new field \cite{Florens07,LeHur&Li05,Mebrahtu12}:
$ \widetilde{\Phi}_{\sigma}(x)=\sqrt{g}(\Phi_{\sigma}(x)+\phi(x))$ with $\quad g=1/(1+2 r)$, which
satisfies $\langle e^{-i\widetilde{\Phi}_{\sigma}(x=0,t)} e^{i\widetilde{\Phi}_{\sigma}(x=0,0)} \rangle\sim t^{-1}$.
Note that only $\phi(x=0)=\phi$ has the physical meaning (i.e. phase fluctuation), and $\phi(x\neq 0)$ are auxiliary fields.
Overall, we have $[\phi(x),\phi(x')]= 2 i r \pi\, \rm{sgn} (x-x')$. Since the tunneling involves only the phase $\phi(x=0)$,
the conductance will not be affected by the auxiliary fields.
Then, the Hamiltonian becomes
\begin{eqnarray}
 H &&= \sum_{\sigma} \frac{v_F}{4\pi} \int_{-\infty}^{\infty} dx \big( \partial_{x}\widetilde{\Phi}_{\sigma}(x) \big)^2 \nonumber\\
  && +\Big[ y_{\rm d, \sigma}  \frac{F_{\sigma}}{\sqrt{2\pi}} e^{-i \widetilde{\Phi}_{\sigma}(0)/\sqrt{g}} \, d 
   + y_{\rm c, \sigma} \frac{F_{\sigma}}{\sqrt{2\pi}} e^{-i \widetilde{\Phi}_{\sigma}(0)/\sqrt{g}} d^{\dagger} \nonumber\\ 
  && + h.c. \Big] 
 +  \frac{K_{\sigma}}{\sqrt{g} \pi}(d^{\dagger}d-1/2)\partial_{x}\widetilde{\Phi}_{\sigma}(0).
 \label{eq:H_LLMF}
\end{eqnarray}
One can define a set of dimensionless parameters: $\widetilde{y}_{d,\sigma}=y_{d,\sigma}l/\sqrt{2\pi}$,
$\widetilde{y}_{c,\sigma}=y_{c,\sigma}l/\sqrt{2\pi}$, and $\widetilde{K}_{\sigma}=2K_{\sigma}/(\pi v_{F})$,
where $l$ is a short time cutoff in the scaling process.
Following the dimension analysis and operator product expansion \cite{CardyBookRG,Senechal,supp}, 
one can simply obtain the RG equations in the weak tunneling limit
\begin{eqnarray}
  \frac{d y_{d,\sigma}}{d\ln l}&=&\Big( 1-\frac{(1-\widetilde{K}_{\sigma})^2}{2g} -\frac{(\widetilde{K}_{-\sigma})^2}{2g} \Big) y_{d,\sigma}, \nonumber\\
  \frac{d y_{c,\sigma}}{d\ln l}&=&\Big( 1-\frac{(1+\widetilde{K}_{\sigma})^2}{2g}-\frac{(\widetilde{K}_{-\sigma})^2}{2g} \Big) y_{c,\sigma}, \nonumber\\
  \frac{d \widetilde{K}_{\sigma}}{d\ln l} &=& 2(1-\widetilde{K}_{\sigma}) \widetilde{y}_{d,\sigma}^2 - 
       2 (1+\widetilde{K}_{\sigma}) \widetilde{y}_{c,\sigma}^2  \nonumber\\
  && - 2 \widetilde{K}_{\sigma} \widetilde{y}_{d,-\sigma}^2 - 2 \widetilde{K}_{\sigma} \widetilde{y}_{c,-\sigma}^2 .
  \label{eq:RGEQ_spinful}
\end{eqnarray}
Five fixed points are obtained and shown in Fig. \ref{fig:impurityBS} (b). 
The first one corresponds to $\widetilde{K}_{\uparrow}=0$, 
$\widetilde{K}_{\downarrow}=-1$, $y_{d,\uparrow}=y_{d,\downarrow}=y_{c,\uparrow}=0$.
In this case, $y_{c,\downarrow}$ will flow to perfect transmission, $d y_{d,\uparrow}/d\ln l=-2r y_{d,\uparrow}$,
$d y_{d,\downarrow}/d\ln l=(-1-4r) y_{d,\downarrow}$, and $d y_{c,\uparrow}/d\ln l=-2r y_{c,\uparrow}$.
The leading tunneling process corresponds to $y_{d,\uparrow} \cdot y_{c,\downarrow}$,
i.e. a spin-up electron entering the ZEIBS from the lead, then hopping out to form a cooper pair with another spin-down
electron from the lead. Therefore, the zero-voltage conductance shows a power law decay $G\sim T^{2r}$ near $T=0$. 
The finite voltage bias will cut off the scaling, and thus the ZBP will split at low $T$.
Conductance shows the same power law decay for three other similar fixed points. Unless the initial condition
$y_{d,\sigma}=y_{c,\sigma}$ is satisfied, the system will flow to one of these four fixed points.

If the bare parameters reach a symmetric point: $K_{\sigma}=0$ and $y_{d,\sigma}=y_{c,\sigma}$, 
all the tunneling strength $y_{d(c),\sigma}$ is relevant and will flow to perfect transmission (i.e. perfect Andreev reflection);
this condition leads to an unstable critical point which belongs to the same universal 
class as the case of tunneling into a MF. 
By noting the similarity between our model (i.e. Eq.~\ref{eq:H_LLMF}) 
and the case with a Luttinger liquid lead \cite{supp}, one can obtain
the $V=0$ conductance for this symmetric point (or for MF) in the strong coupling limit (low $T$) 
\cite{supp,Fidkowski12}:  $2e^2/h-G\sim T^{(2-4r)/(1+2r)}$.
For ZEIBS, the condition  $y_{d,\sigma}=y_{c,\sigma}$ requires fine tuning both the tunneling barrier and spin components,
and thus its realization is extremely difficult.

\textit{Tunneling into a cluster of mid-gap states} --- 
If both the spin rotation and time reversal symmetries are broken in
SCNW, disorder can induce a cluster of mid-gap states around zero energy localized near
the end of the wire \cite{Bagrets12,Liujie12,Neven13}. 
Therefore, even without a zero energy state (either MF or ZEIBS),
the tunneling conductance shows a zero-energy peak at finite $T$ without dissipation effect. 
To study the dissipation effects for those cases,
we consider the tunneling Hamiltonian in Eq. (\ref{eq:tunnelingH}), 
and treat the tunneling strength $y$ as a small parameter
such that the perturbation theory can be applied.
This assumption is valid for tunneling into any non-MF state (with a small bare tunneling strength) 
except at the highly symmetric situation 
shown in the previous section. 
 
\begin{figure}[t]
\centering
\vspace{0.1in} 
\includegraphics[width=3.4in,clip]{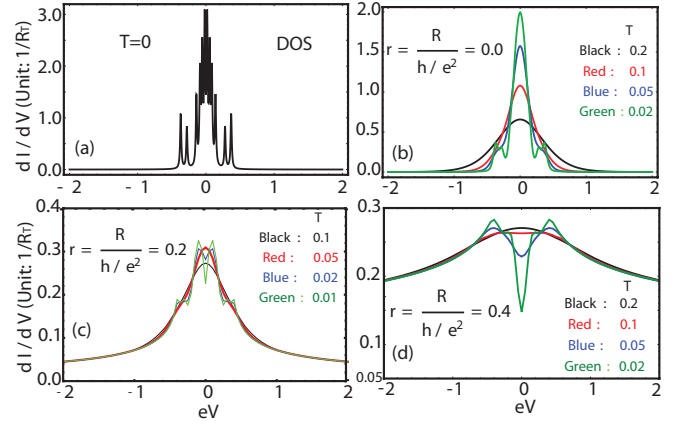}
\vspace{-0.25in} 
\caption{(color online) Differential conductance $dI/dV$ (tunneling into a cluster of mid-gap states around zero energy) 
as a function of applied voltage $V$. (a) An arbitrary choice of the DOS for a cluster of states, which is also
the $T=0$ conductance for $r=0$. (b) The $r=0$ finite temperature conductance. 
The conductance with dissipation effect, i.e. $r=0.2$ (c) and $r=0.4$ (d), for different temperatures.
The single peak splits into two as $T$ decreases.
} 
\vspace{-0.25in} 
\label{fig:cluster_tot}
\end{figure}

The current operator for the junction is 
$\hat{I}=i[H_T,\sum_{k\sigma}  c^{\dagger}_{k\sigma} c_{k\sigma}]=-i\sum_{k\sigma,\nu}(y_{k\sigma,\nu}c^{\dagger}_{k\sigma} b_{\nu} e^{-i \phi} - h.c.)$
Then, the current through the junction up to the leading order in tunneling strength is given by Kubo formula
(this can also be obtained by golden rule \cite{Nazarov&Ingold}) 
\begin{eqnarray}
  I (t) &=& -i \int_{-\infty}^{\infty} dt'\, \theta(t-t')\; \langle [\hat{I}(t), H_T(t')] \rangle_0 \nonumber\\
   &=& \int_{-\infty}^{\infty} \frac{d\omega_1}{2\pi} \int_{-\infty}^{\infty} \frac{d\omega_2}{2\pi} 
      \sum_{k\sigma,\nu} |y_{k\sigma,\nu}|^2 A_{k\sigma}^L(\omega_1)  A_{\nu}^{SCNW}(\omega_2) \nonumber\\
   &&\times \{ [1-f(\omega_1 - eV)]f(\omega_2)P(\omega_2 - \omega_1)\nonumber\\
   &&- f(\omega_1-eV)[1-f(\omega_2)]P(\omega_1 - \omega_2)   \} .
\end{eqnarray}
with
\begin{equation}
 P(\omega)=\frac{1}{2\pi}\int_{-\infty}^{\infty} dt \exp[i\omega t + J(t)]
\end{equation}
where $J(t) = \langle \phi(t) \phi(0) \rangle -\langle \phi^2 \rangle$ 
(see \cite{Nazarov&Ingold,supp} for more details)
and $\langle\cdots\rangle_0$ indicates 
the average without the tunneling term. $P(\omega)$ describes the energy emission and absorption
in the electron tunneling processes due to dissipation effects.
$A_{k\sigma}^L(\omega_1)$ is the spectral function of the lead, 
and we assume a constant density of state (DOS): 
$\sum_{k\sigma}|y_{k\sigma,\nu}|^2 A_{k\sigma}^L(\omega_1)=1/(e R_T)$,
where $R_T$ can be viewed as the tunneling resistance. $f$ is the Fermi-distribution function. 
Without dissipation, i.e. $r=0$, at zero temperature one obtain 
$dI/dV\propto \sum_{\nu} A_{\nu}^{SCNW}(\omega_2)$ which
gives the DOS of the wire. A realization of the DOS (i.e. $T=0$ conductance for $r=0$), 
is shown in Fig.~\ref{fig:cluster_tot} (a). For finite temperature, 
this cluster of states results in a ZBP as shown in Fig.~\ref{fig:cluster_tot} (b). 
As temperature decreases (still larger than the level spacing of the mid-gap states), the ZBP height increases for $r=0$,
which is similar to Majorana ZBP. This feature changes dramatically when the dissipation effect is included.
As shown in Fig.~\ref{fig:cluster_tot} (c) $r=0.2$ and (d) $r=0.4$ ($R\sim k\Omega$), 
the single conductance peak splits into two peaks 
and zero bias conductance decreases as temperature goes down; 
and this feature is contrary to that of Majorana ZBP : The zero bias conductance for $r<1/2$
increases as $T$ goes down and finally approaches $2e^2/h$ at $T=0$.
Fig.~\ref{fig:cluster_tot} (c) and (d) also show that the peak splitting occurs at higher $T$ 
for larger $r$.

\textit{Discussion} ---
Tunneling into a MF is equivalent to the resonant tunneling between 
an electron lead and a hole lead \cite{law09} (also see Eq.~(\ref{eq:tunneling_MF}))
with exactly the symmetric tunneling barriers due to the topological properities of MF.
With ohmic dissipation, the resonant tunneling shows non-trivial 
phase diagrams \cite{Mebrahtu12,LiuDRL13}: 1) any asymmetry in the barriers induces
a relevant backscattering which destroys the resonant tunneling; 2) this backscattering
vanishes for a special symmetric point, and the next leading term is irrelevant for small $r$
($r<1/2$ for our case). This symmetry, which results in dissipative resonant tunneling,
is topologically protected by 
MF; it is not protected for other cases, and requires fine tuning.
In the experiments \cite{kouwenhovenSCI12,deng12,das12}, the metal lead can be made
rather resistive ($R\sim k\Omega$, but need $r<1/2$), by using e.g. $\rm{Cr/Au}$ 
film \cite{Mebrahtu12,Mebrahtu13}.
When coupling to a MF zero mode, the height of ZBP
increases as $T$ goes down: $2e^2/h-G\sim T^{(2-4r)/(1+2r)}$ near $T=0$, and
$G\sim T^{2r-1}$ for high $T$.
When coupling to a non-MF mode causing a ZBP, however, its height
shows a power law suppression at low $T$: $G\sim T^{2r}$.

D.E.L. is grateful to H.U.Baranger and A. Levchenko for valuable discussions
and suggestions. 
The author acknowledges support from US DOE, Division of Materials
Sciences and Engineering, under Grant No.\,DE-SC0005237, Michigan state university, and
ARO through contract W911NF-12-1-023.

\bibliography{DissipativeMF}

\end{document}